\documentclass[twocolumn,preprintnumbers,amsmath,amssymb]{revtex4}
\usepackage[dvipdfmx]{graphicx}
\usepackage{dcolumn}
\usepackage{bm}
\usepackage{color}
\usepackage{ulem}
\usepackage{pdfpages}
\setboolean{@twoside}{false}

\begin{document}
\title{Colossal Seebeck coefficient of hopping electrons in (TMTSF)$_{2}$PF$_{6}$}
\author{Yo Machida$^{1}$, Xiao Lin$^{2}$, Woun Kang$^{3}$, Koichi Izawa$^{1}$ and Kamran Behnia$^{2}$\email{kamran.behnia@espci.fr}}
\affiliation{$^{1}$Department of Physics, Tokyo Institute of Technology, Meguro 152-8551, Japan\\
$^{2}$LPEM (UPMC-CNRS) ESPCI, 75005 Paris, France\\
$^{3}$Department of Physics, Ewha Womans University, Seoul 120-750, Korea}

\date{February 22, 2016}

\begin{abstract}
 We report on a study of Seebeck coefficient and resistivity in the quasi-one-dimensional conductor (TMTSF)$_{2}$PF$_{6}$ extended deep into the Spin-Density-Wave(SDW) state. The metal-insulator transition at $T_{SDW}$ = 12 K leads to a reduction in carrier concentration by seven orders of magnitude. Below 1 K, charge transport displays the behavior known as Variable Range Hopping (VRH). Until now, the  Seebeck response of electrons in this regime has been barely explored and even less understood. We find that in this system, residual carriers, hopping from one trap to another, generate a Seebeck coefficient as large as 400 $k_{B}$/$e$. The results provide the first solid evidence for a long-standing prediction according to which hopping electrons in presence of Coulomb interaction can generate a sizeable Seebeck coefficient in the zero-temperature limit.
\end{abstract}
\maketitle

The Seebeck coefficient, a measure of entropy per mobile particle \cite{callen1948,macdonald1962,ziman1964,behnia2015}, behaves differently in metals and insulators. In a Fermi-Dirac distribution, entropy is confined to an energy window centered at the Fermi energy with a width of $k_{B}T$. A fixed population of mobile electrons share this shrinking entropy when a metal is cooled down. Therefore, the diffusive Seebeck coefficient of a metal, below the degeneracy temperature, has an upper limit of $\frac{\pi^{2}}{3}\frac{k_{B}}{e}\sim$ 288 $\mu$V/K and its linear decrease with temperature reflects the quadratic temperature dependence of the energy shift between the chemical potential and the Fermi energy\cite{ziman1964}. In an insulator, both entropy and mobile carriers vanish at zero temperature and the fate of the Seebeck coefficient depends on the relative rate of decrease in these two vanishing quantities. Since there is a well-defined energy gap, $\Delta$, between the chemical potential and the nearest occupied energy level, the Peltier coefficient would be of the order of $\Delta$ and the Kelvin relation implies  the Seebeck coefficient to be proportional to the inverse of absolute temperature \cite{macdonald1962}(See Fig. 1).

\begin{figure}
\includegraphics[width=8cm]{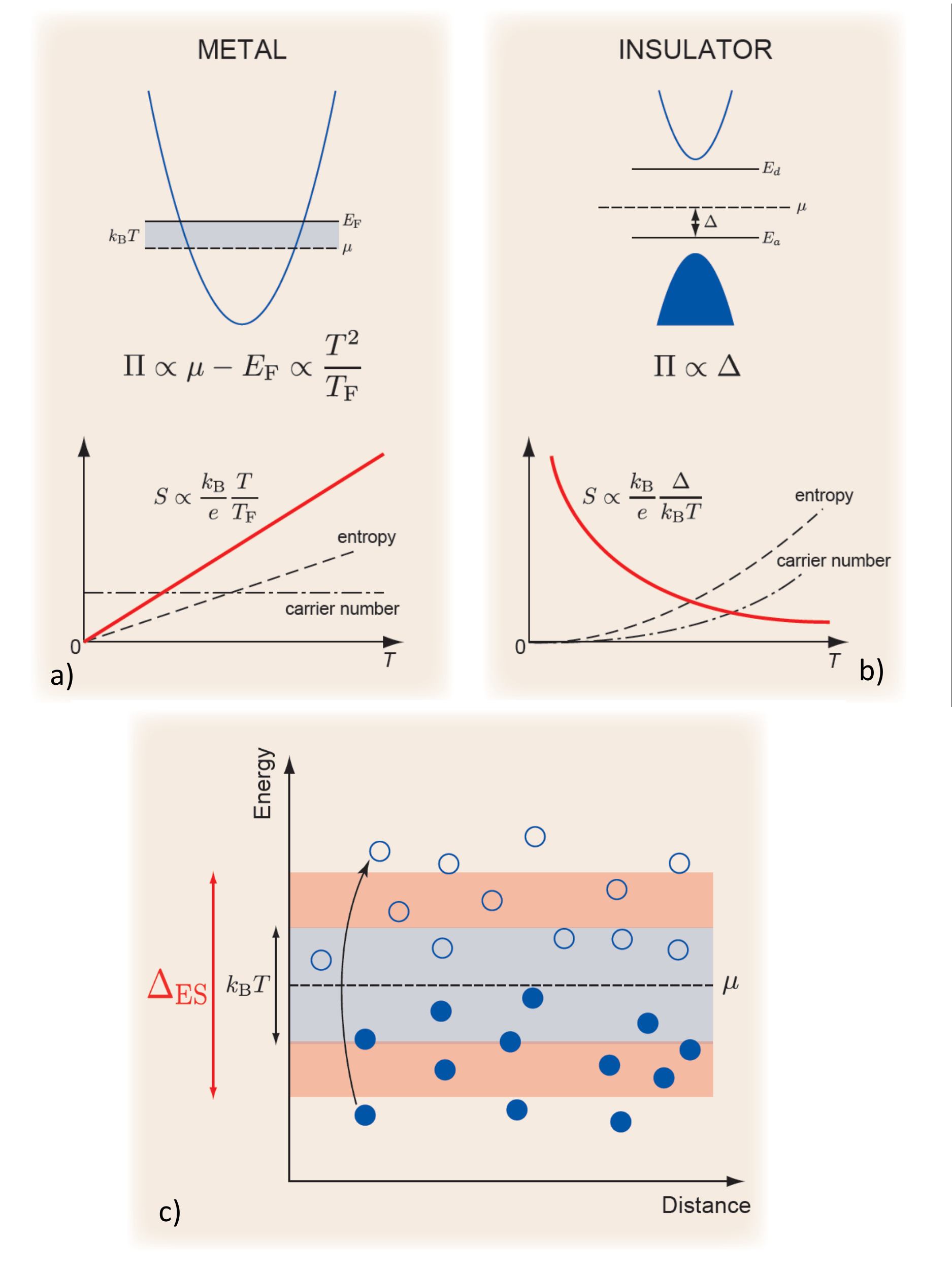}
\caption{ a) In a metal, the average thermal energy of a carrier, which sets the Peltier coefficient, $\Pi$, is quadratic in temperature. With decreasing temperature, mobile entropy shrinks but carrier number does not change. As a consequence of Kevin relation ($\Pi =S T$), the Seebeck coefficient is T-linear. b) In an insulator , the average thermal energy of a carrier is T-independent, set by the distance between the closest occupied level and the chemical potential. Both the carrier number and entropy vanish at zero temperature. If the carrier number decreases faster, the Seebeck coefficient will increase with decreasing temperature. c) Real insulators will eventually enter the Variable Range Hopping regime, where Coulomb interaction opens a soft gap ($\Delta_{ES}$) in the vicinity of the chemical potential. What happens to the thermoelectric response at temperatures below this gap?}
\end{figure}

Any real insulator cooled down towards zero temperature, however, would end up entering a regime in which electronic transport is governed by carriers trapped in local defects and jumping from one site to another, a regime dubbed Variable Range Hopping (VRH) (See Fig.1c). Would this impede the survival of a finite Seebeck coefficient in the zero-temperature limit? In spite of several theoretical proposals addressing this question\cite{zvyagin1973,whall1981,burns1985,lien1999,mahan2015}, no satisfactory response has been given to this question. There is not theoretical consensus, as theorists have variously predicted that in insulators cooled down to the lowest achievable temperature, one is expected to see a vanishing\cite{zvyagin1973}, a finite\cite{burns1985}, or a diverging [but unmeasurable]\cite{mahan2015} Seebeck coefficient. On the experimental side, there is no track of a result providing a definite answer to this question. A large Seebeck coefficient was found in early experiments on semiconducting silicon and germanium\cite{geballe1954,geballe1955}, but the data acquisition was interrupted at a temperature too high to resolve the asymptomatic response in the zero-temperature limit.

In this paper, we present a study of electric resistivity and Seebeck coefficient in (TMTSF)$_{2}$PF$_{6}$, a quasi-one-dimensional conductor, known to go through a nesting-driven Spin-Density-Wave instability(SDW) at $T_{SDW}$ = 12 K. As the first pressure-induced organic superconductor\cite{jerome1980}, this Bechgaard salt has been subject to numerous studies during more than three decades(See \cite{jerome2012,brown2015} for recent reviews).  According to our findings, below 1 K, electric resistivity displays a VRH temperature dependence and concomitantly the Seebeck coefficient rapidly increases with decreasing temperature, attaining a magnitude as large as 37 mV/K at $T \sim$ 0.1 K. A quantitative description of our thermoelectric data is missing. Nevertheless, this is the first explicit experimental confirmation of the persistence of a finite Seebeck coefficient in an insulating solid in the zero-temperature limit. We argue that this arises as a consequence of the huge number of configurations available to a hopping electron. Our result is in qualitative agreement with those subset of theoretical proposals\cite{burns1985,lien1999,mahan2015}, which do not predict a vanishing fate for the Seebeck coefficient in a zero-temperature insulator.

\begin{figure}
\includegraphics[width=9cm]{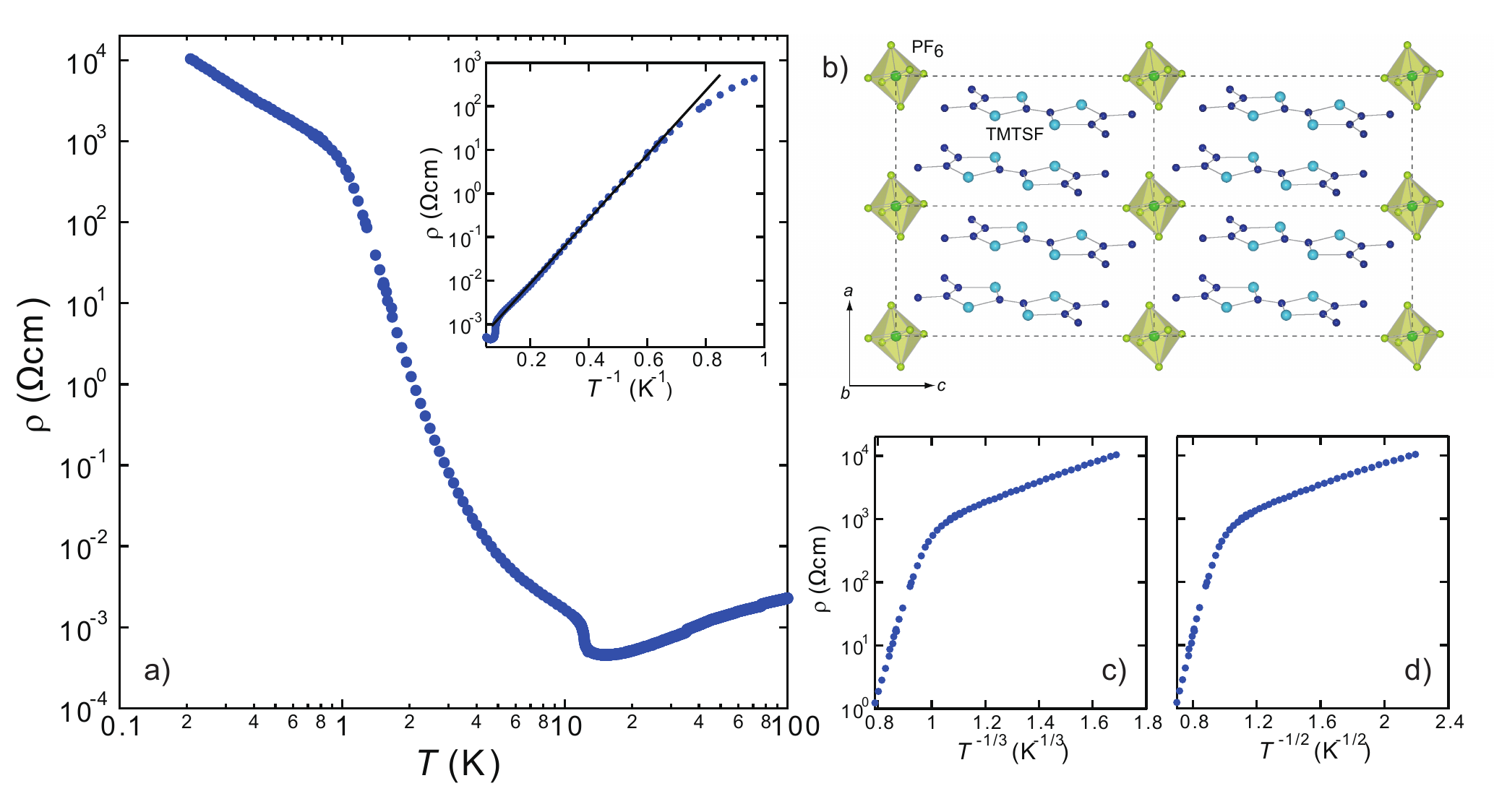}
\caption{ a) Temperature dependence of resistivity in a logarithmic plot; Resistivity increases by more than seven orders of magnitude. As seen in the inset, over a wide temperature window (10 K to 1 K), it follows an activated behavior. b) Crystal structure of the Bechgaard salt (TMTSF)$_{2}$PF$_{6}$. c) and d) Below 1 K, resistivity displays a exp[($T/T_{0}$)$^{-\gamma}$]  behavior characteristic of VRH. Semi-logarithmical plots of $\rho$ vs. $T^{-1/3}$(c) or $T^{-1/2}$(d) both yield quasi-straight lines at low temperature end.}
\end{figure}

Fig.~2a shows the temperature dependence of the resistivity. As found in previous studies \cite{chaikin1981,kim1995,petukhov2005,gruner1994}, the SDW transition drastically affects resistivity. Cooling the sample down to 0.2 K leads to a seven-order-of-magnitude enhancement in resistivity. Between 10 K and 1 K, it follows an activated behavior with a temperature dependence expressed as $\rho \propto {\rm exp}(\Delta/k_{B}T)$ and the extracted $\Delta$ of 20 K is comparable with what was previously reported and what is expected for a mean-field transition occurring at 12 K. Below 1 K,  we resolve a clear downward deviation from the activated behavior. Note that the million-fold increase in resistivity  indicates that the carrier number has dropped to a level that there remains one mobile electron per 10$^{6}$ TMTSF. This puts an upper limit to the number of surviving residual carriers after cooling down to this temperature. This is an insulator, in the strict sense of the term, a solid   with divergent unsaturated resistivity lacking mobile electrons at zero temperature.

One source of residual carriers at finite temperature are crystal defects,  potential wells holding trapped charge carriers. Electric conductivity in this context is described along the lines first drawn by Mott\cite{mott1968} and dubbed Variable Range Hopping\cite{reviewVRH,adam2014}. In agreement with a previous study \cite{kim1995}, we find that resistivity below 1 K can be described by the expression $\rho \propto {\rm exp}[(T/T_{0})^{-\gamma}]$. However, our results cannot pin down the magnitude of $\gamma$ which can be a number between 1/3 and 1/2 as seen in Fig.~2c and 2d.

\begin{figure}
\includegraphics[width=7cm]{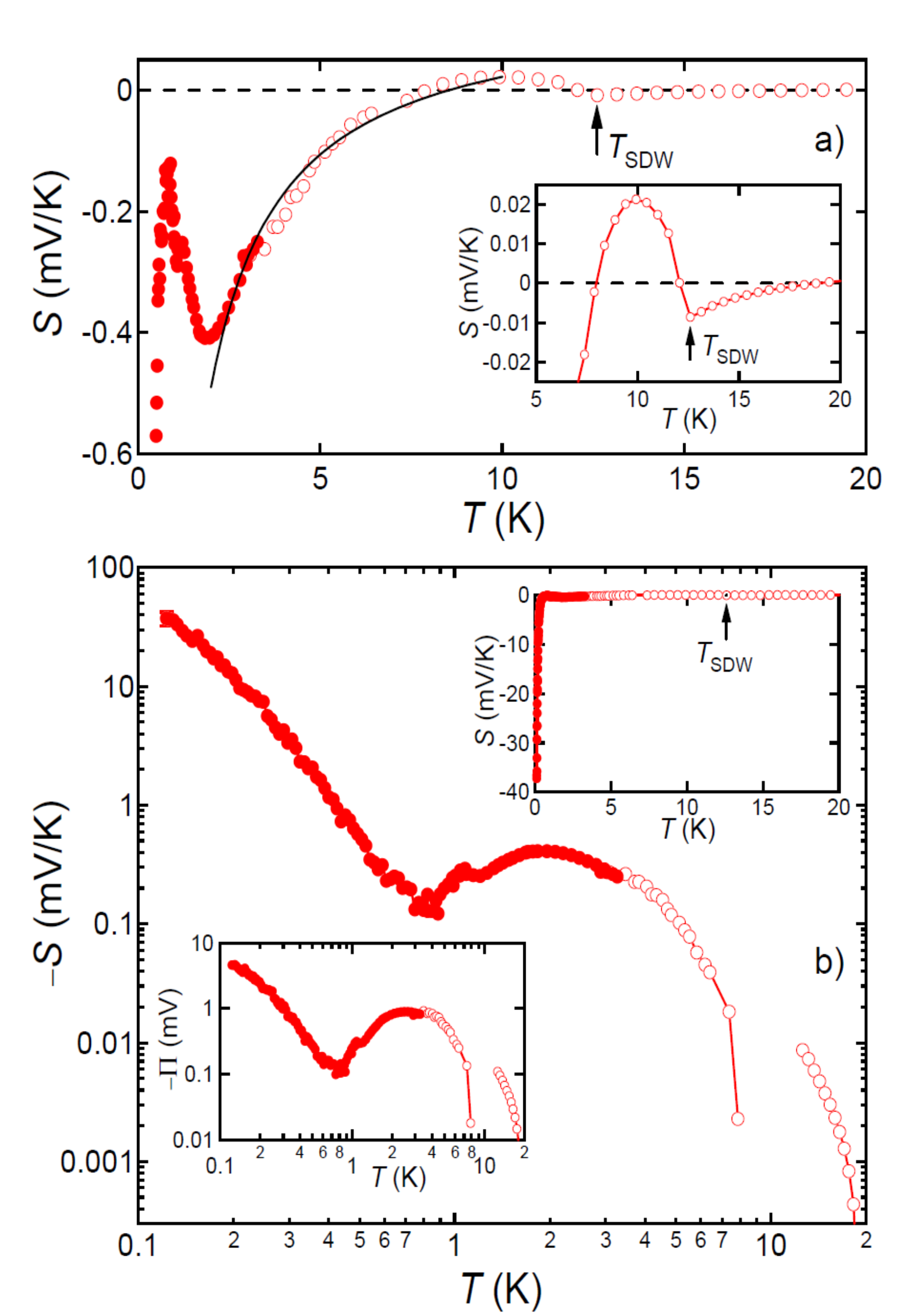}
\caption { a) Temperature dependence of the Seebeck coefficient in the vicinity of the SDW transition. Below the transition temperature, the Seebeck coefficient follows a $T^{-1}$ dependence as indicated by a solid line. The inset shows the anomaly at the SDW transition. b) Temperature dependence of the Seebeck coefficient in a logarithmic plot. The magnitude of the Seebeck coefficient attains 37 mV/K at 0.13 K. The lower left inset shows the temperature dependence of the Peltier coefficient $\Pi=ST$.  The upper right inset shows the data in a linear plot. Solid and empty circles represent data obtained on two different samples with two different set-ups.}
\end{figure}

We now turn our attention to the Seebeck coefficient and its temperature dependence illustrated in Fig.~3. As seen in the inset of Fig.~3a, the SDW transition leads to a jump in the Seebeck coefficient, as reported by previous studies\cite{mortensen1982,choi1985, chai2007}. The magnitude of the positive jump in $S$ seen here is larger than what was reported by previous works (See the supplement). It is concomitant with a sharp jump in resistivity and a lambda-anomaly in specific heat \cite{coroneus1993,powell2001}, all three confined to a very narrow window near the critical temperature.

\begin{figure}
\includegraphics[width=8cm]{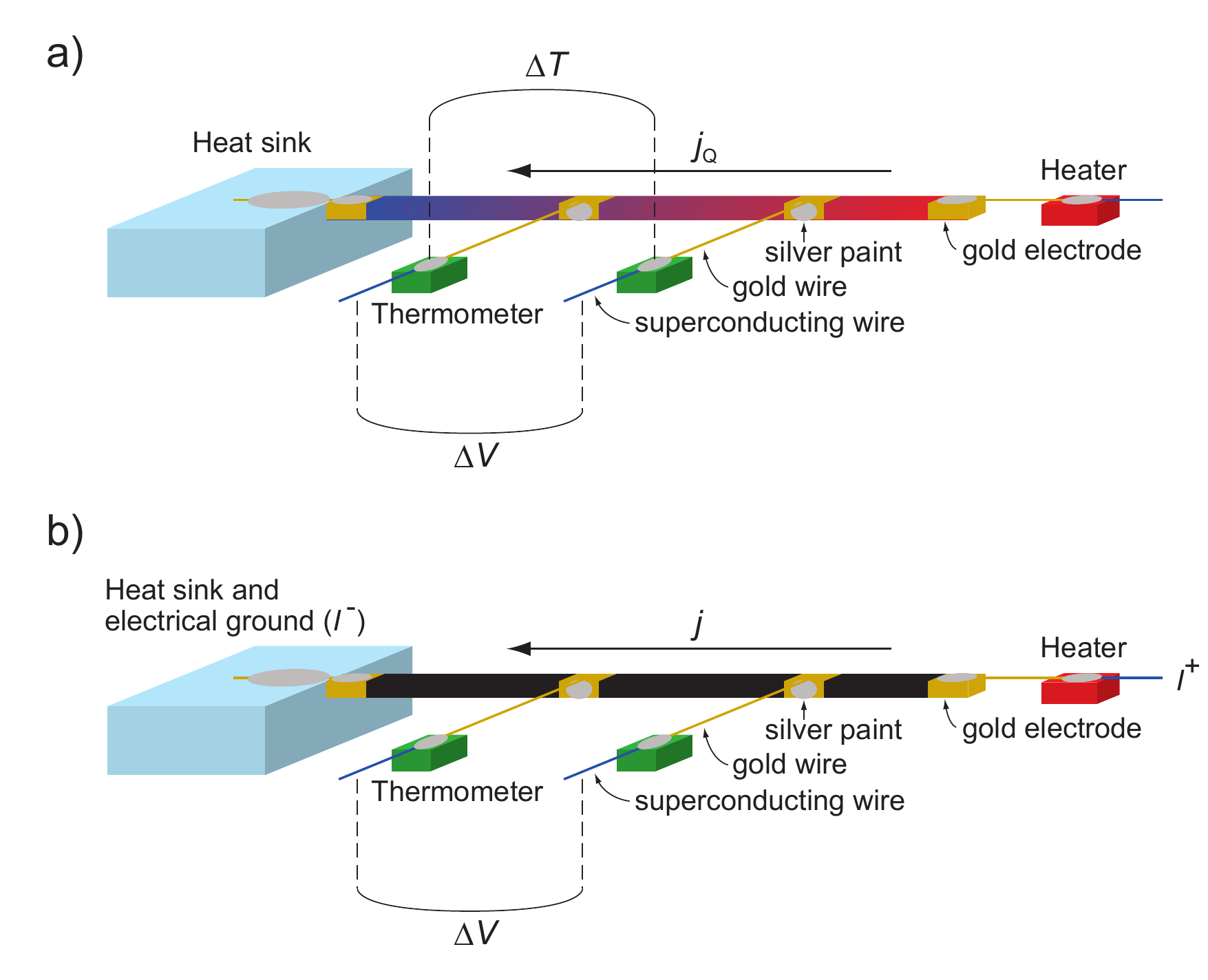}
\caption{ a) For measuring the Seebeck coefficient, a heat current was injected to the sample by passing a current though the heater and measuring both the temperature gradient and  the voltage difference created by this heat current. b) For measuring the electrical resistivity, a current was injected in to the sample and the voltage difference created by this charge current. In both cases, the voltage was measured in the same way by the same electrodes and the same instrument.}
\end{figure}

Below the SDW transition temperature, the Seebeck coefficient follows an activated behavior $|S|\propto\Delta/k_BT$ with $\Delta\sim$ 20-30 K, which is comparable with the one extracted from resistivity, and peaks to approximately 0.5 mV/K at $T$ $\sim$ 2 K. This is concomitant with a peak in the Peltier coefficient (extracted using the Kelvin relation) of 1 mV (see the lower inset of Fig.~3b), quantifying the average thermal energy carried across the gap.  A large Seebeck coefficient with a Seebeck coefficient roughly proportional to the inverse of temperature has been seen in several quasi-one-dimensional organic conductors upon the entry of the system in the SDW state\cite{mortensen1983,mortensen1984}.  A new and unexpected behavior is detected upon further cooling.  Below a local peak  at $T$ $\sim$ 2 K, a spectacular enhancement in the Seebeck coefficient, with S $\propto T^{-2.5}$ is detectable as soon as the system enters the VRH regime below 1 K. This result, a Seebeck coefficient  attaining a  magnitude as large as 37 mV/K, almost 400 times $k_{B}/e$ is the main new result of the present paper. It is highlighted in the upper inset of Fig.~3b by presenting a linear plot of the data. The enhancement below 1 K easily dwarfs the anomaly seen at the SDW transition. The lower inset of Fig.~3b presents the temperature dependence of the Peltier coefficient. Its low-temperature magnitude implies that hopping electrons carry an average energy of 4 meV at 0.13 K. Let us examine the possible origins of this colossal thermoelectric response.

Phonon drag can lead to a large amplification of the Seebeck coefficient\cite{macdonald1962,behnia2015} in cryogenic temperatures. This happens when thermally-induced flow of phonons pull electrons along their way giving rise to a finite electric field. As recently argued\cite{mahan2015}, it cannot paly a major role here. A large phonon drag contribution suddenly setting in below 1 K is implausible, since the electron-phonon coupling is weak and lattice thermal conductivity rapidly decreasing.

Can the result be an experimental artefact caused by the large resistance of the sample? This is also unlikely. We measured three different samples which yielded comparable results. Moreover, as seen in the figure, there is a satisfactory match between the low-temperature data (obtained in Tokyo on one sample) and the high-temperature data (acquired in Paris on another sample). Our standard one-heater-two-thermometer set-up is illustrated in Fig.4. A similar setup has  been used to measure the sub-Kelvin thermoelectric response in a variety of solids including heavy-electron metals\cite{izawa2007,machida2012}, superconducting thin films\cite{pourret2006} and semimetals\cite{zhu2010}. An additional complication in our case is the sample resistance, as large as 10 M$\Omega$ at the lowest temperature. However, the input impedance of the measuring instrument was two orders of magnitude larger than this. The same instrument was used to measure the voltage in our resistivity measurement, with reliable results. We cannot think of any experimental artefact leading to the large ratio of voltage to thermal gradient observed here(See the supplement for more details).

As seen above, the drastic enhancement in the Seebeck coefficient is concomitant with the entry to the VRH transport regime detected by resistivity. It is therefore natural to look for an additional source of thermoelectric response associated with this regime.  Mott\cite{mott1968} argued that when carriers hop from one site to the other with a probability proportional to exp($-W/k_{B}T-2R/\xi$) ($W$ is the energy separation and $R$ is the spatial distance between the two sites, while $\xi$ is the localization length of the hopping electron), the expression for electric conductivity in a system with dimension $d$ becomes:

\begin{equation}
\sigma\propto \exp[-(T_{0}/T)^{1/d+1}]
\end{equation}

Efros and Shklovskii showed that a finite Coulomb interaction leads to a significant depopulation of the occupied sites in the immediate vicinity of the chemical potential and the opening of a soft gap, $\Delta_{ES}$ \cite{efros1975}. The resulting expression for electric conductivity is independent of dimensionality :
\begin{equation}
\sigma\propto \exp[-(T'_{0}/T)^{1/2}]
\end{equation}

The presence of this Coulomb gap is expected to become visible when the temperature is low enough such that $k_{B}T<\Delta_{ES}$. In practice, as our present data shows (See Fig.~2c and 2d), it is hard to distinguish between the two kinds of stretched exponentials. Burns and Chaikin argued  that the fate of the thermopower in the zero-temperature limit is drastically modified by the presence of the Coulomb gap\cite{burns1985}. In the absence of the Coulomb gap,  thermopower should be vanishing:
\begin{equation}
S(T)\mid_{VRH} \propto T^{d-1/d+1}
\end{equation}

In three dimensions ($d$ = 3), this expression is identical to the one first found by Zvyagin \cite{zvyagin1973} ($S(T) \propto (TT_{0})^{1/2}$). On the other hand, Burns and Chaikin argued that in presence of Coulomb gap, thermopower will remain finite in the low-temperature limit:
\begin{equation}
S(T)\mid_{VRH}^{ES} \propto S_0
\end{equation}

Later and with detailed calculations, Lien and Toi confirmed this conclusion\cite{lien1999}. However, this theoretical prediction has never been confirmed by experiment. To put it in a few words: Can the Seebeck coefficient of a solid remain finite in the zero-temperature limit, the third law of thermodynamics notwithstanding? To the best of our knowledge, this fundamental question is answered for the first time by the observation reported here.

Many questions remain unanswered. Is this result generic to all semiconductors cooled below their Coulomb gap in the VRH regime?  Available reports on thermoelectric response in archetypal semiconductors such as silicon\cite{geballe1955,weber1991} and germanium\cite{fredrikse1953,geballe1954,goff1965} resolve a large Seebeck coefficient down to the lowest temperature of measurement. But the data stops above 5 K. Only in metallic samples of silicon\cite{ziegler1996} or germanium\cite{goff1965} (that is, with carrier density above the threshold of the metal-insulator transition), a vanishing thermopower in the low-temperature limit has been resolved. Therefore, the thermoelectric response of band insulators in the VRH limit remains an open question\cite{supplement}. Here, the robust insulating ground state owes its existence not to a band gap,  but to a many-body gap opened by a SDW transition. Does this matter? This is another open question. Because of the incommensurability of the SDW order in (TMTSF)$_{2}$PF$_{6}$, one expects the presence of collective excitations known as phasons deep inside the ordered state\cite{barthel1993}. We note also that another unsolved enigma is the electric-field dependence of non-linear conductivity  below 1 K \cite{gruner1994}. One may wonder about the relevance of these other puzzling features of (TMTSF)$_{2}$PF$_{6}$ to the observation reported here.

If a large Seebeck coefficient happens to be a generic feature of an insulator cooled below the temperature corresponding to its Coulomb gap, then a new research avenue opens up. How to quantify the Seebeck coefficient of hopping electrons in each system? The Heikes formula\cite{heikes1961} is often used to describe the magnitude of the thermoelectric response by hopping electrons. As Mott noted long ago\cite{mott1968}, this formula assumes a single hopping probability for all sites and therefore, it cannot be readily used in the VRH regime. The connection between the energy landscape carved by defects in a real insulator and the low-temperature thermoelectric response is an unexplored field of research. The configurational entropy of each hopping electron is the information which travels with it. Therefore, there are potential links to the emerging field of thermodynamics of information\cite{parrondo}.

In summary, we measured the Seebeck coefficient (TMTSF)$_{2}$PF$_{6}$ down to very low temperatures and found that the entry to the VRH regime is concomitant with a very large enhancement of the Seebeck coefficient, which attains a magnitude as  large as 37 mV/K implying that a finite Seebeck coefficient can persist in a solid in the zero-temperature limit.

This work was supported in Japan by JSPS KAKENHI Grant Number 25400361, 23340099 and 15K05884 (J-Physics), in France by ANR through the SUPERFIELD project and in Korea by the NRF grants funded by the Korea Government (MSIP) (No. 2015-001948 and No. 2010-00453).

\newpage

\appendix

\section{Experimental methods}
\begin{figure}
\includegraphics[width=8cm]{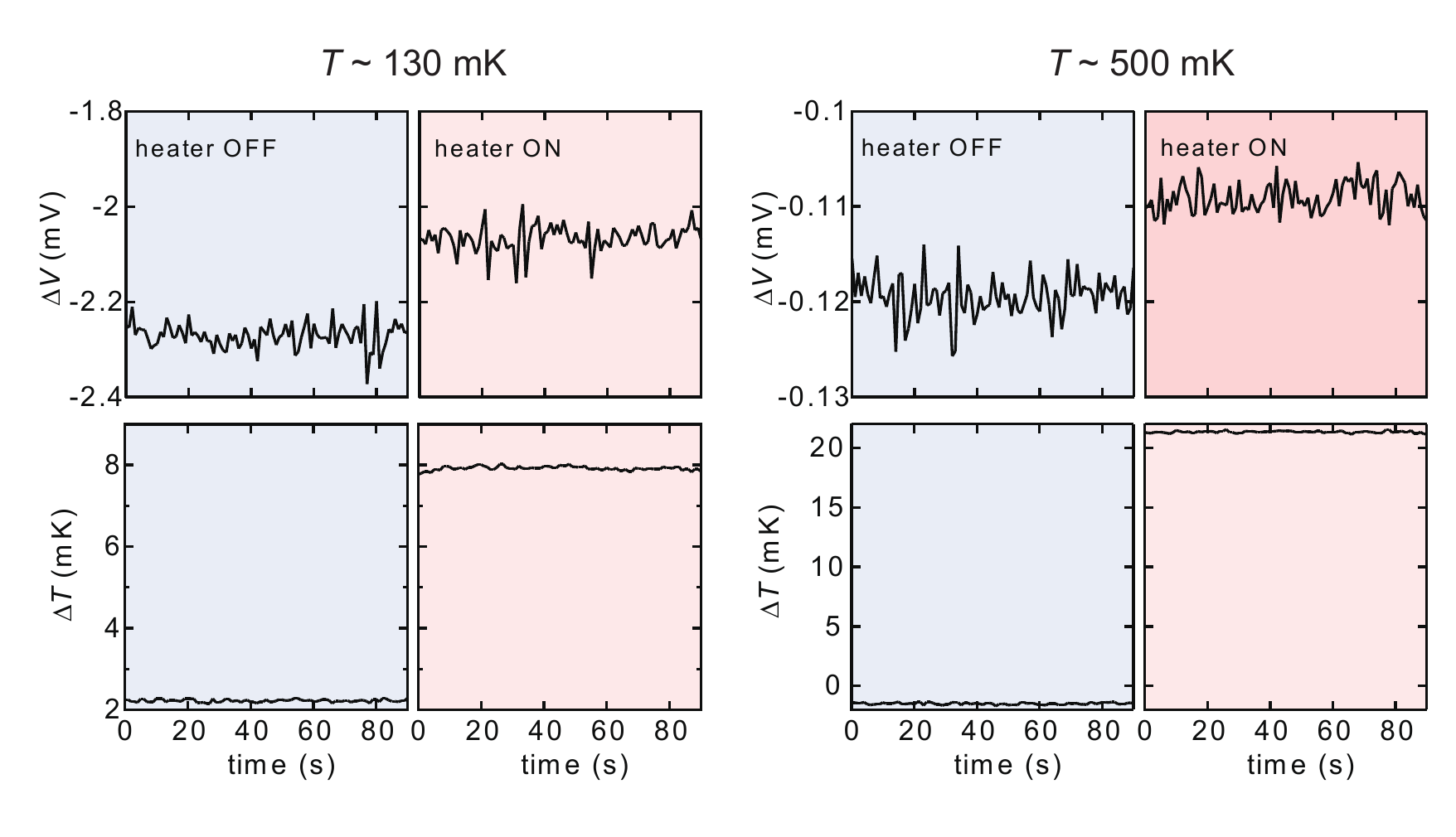}
\
\begin{flushleft}
{\bf Figure S1} The measured thermal gradient and thermoelectric voltage in absence and in presence of a heat current at two different temperatures. As the tempertaure decreases the resistance of the sample increases making the output voltage noisier. In spite of the large noise level, a clear shift in the voltage can be clearly resolved at both temperatures.
\end{flushleft}
\end{figure}

Single crystals of (TMTSF)$_2$PF$_6$ with dimensions of approximately 2.0 $\times$ 0.1 $\times$ 0.1 mm$^3$ were grown by a conventional electrochemical technique from commercially available products in 1,2-dichloroethane electrolyte solutions.

The Seebeck coefficient was measured using a steady-state two-thermometers, one-heater technique. A thermal gradient was applied along the sample by heating a chip resistor with the Joule effect. The thermal gradient $\Delta T=T_1-T_2$ was determined by measuring the local temperature with two thermoemeters (Cernox resistors in the $^4$He cryostat and RuO$_2$ resistors in the $^3$He-$^4$He dilution refrigerator).  The thermoelectric voltage $\Delta V$ was measured by a digital nanovoltmeter. The experimental uncertainties arising from the measurements of the thermoelectric voltage and the thermometer resistances are about 10 $\%$ in the Seebeck coefficient at the lowest temperature and less than 1 $\%$ at 1 K.

Two samples were measured in a $^3$He-$^4$He dilution refrigerator from 4 K down to  0.1-0.2 K (in Tokyo). Another sample was measured in a $^4$He cryostat from 30 K down to 2 K (in Paris). A satisfactory match was seen between different sets of data in the range they overlapped. In order to prevent cracks during the cooling process, the sample was slowly cooled with rates of $\sim$ 0.1-0.3 K/min. The heat current $j_Q$ was applied along the $a$-axis, the highest conductivity direction of the system.

A stability of better than 0.001 $\%$ of both sample temperature readouts was achieved in the entire measurement range. For each stabilized bath temperature, a temperature difference of the order of  1 to 5 $\%$ of the average temperature was applied, with $\bar{T}=(T{_1}-T{_2})/2$ being the mean temperature.

To achieve good thermal contacts, the thermometers, the heater and the cold finger were connected by gold wires and silver paint to gold electrodes
evaporated on the surface of the sample. The contact resistances were a few ohms. The same contacts and gold wires were used for the electrical connection for the measurement of thermoelectric voltage. The gold wires were switched to the superconducting wires on the thermometry for thermal isolation of the sample. The thermometer and heater chips were suspended by polyimide tubes and electrical connections were made with manganin wires. The heat loss along the wires and suspensions for the thermometry is estimated to be a factor of 1000 smaller than the heat flow through the sample.

In general,  the measured thermoelectric voltage is the difference of the thermoelectric power of two materials in contact, namely gold and (TMTSF)$_2$PF$_6$. In this case, the fact that the Seebeck coefficient is two to six orders of magnitude smaller in gold than in (TMTSF)$_2$PF$_6$ allows us to neglect the contribution of gold and the measured $S$ is approximated to be the absolute Seebeck coefficient of (TMTSF)$_2$PF$_6$.

The same setup was used to measure the Seebeck coefficient of the heavy-Fermion compound YbRh$_2$Si$_2$ using a comparable size of crystal in the same temperature range [S1]. The results are reasonably consistent with those obtained by the different group utilizing the different setup [S2].

The same contacts and wires were used for the electrical resistivity measurements by a four-point technique (See Fig. 4). A DC current $j$ was applied along the $a$-axis of the sample using the superconducting wire attached on the heater and the same nanovoltmeter with an input impedance of 10 G$\Omega$, well above the resistance of the sample even at the lowest temperature ($R \sim$ 10 M$\Omega$), was employed for the measurement of electrical voltage. As seen in Fig. S1, at low temperatures, in spite of the large resistance of the sample, the signal to noise ratio is large enough to allow a clear resolution of the voltage produced by the thermal gradient. We also checked that we remain in the linear response regime, by applying different temperature gradients and finding that the voltage-to-temperature-gradient ratio at a given temperature is insensitive to the magnitude of the thermal gradient.

\section{Reproducibility}
\begin{figure}
\includegraphics[width=8cm]{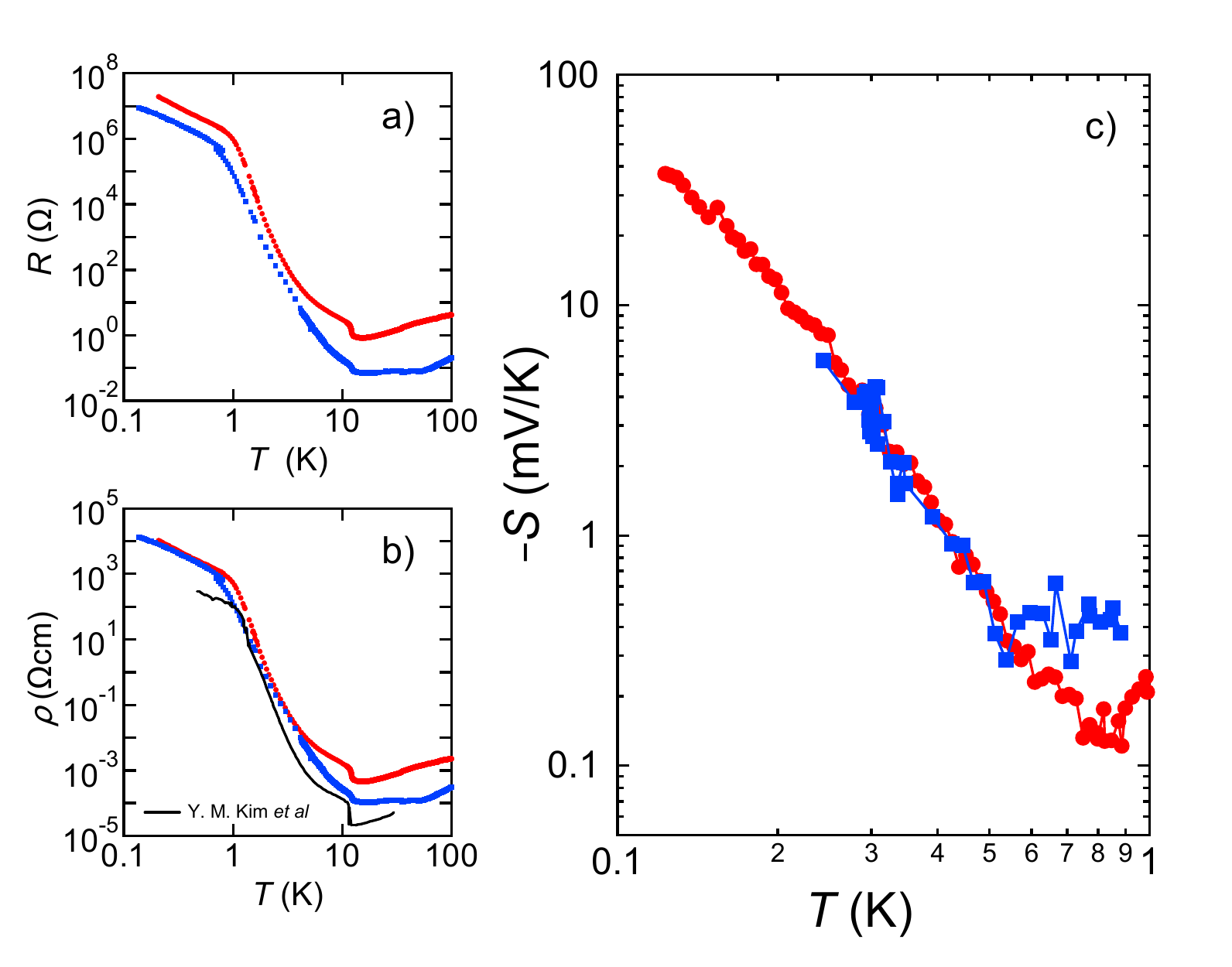}
\
\begin{flushleft}
{\bf Figure S2} Temperature dependence of a) the resistance, b) the resistivity and c) the Seebeck coefficient $S$ for two different samples.
The data for sample no. 1 and no. 2 are represented by red circles and blue squares, respectively.
\end{flushleft}
\end{figure}
To check the reproducibility of our low-temperature data, two different samples were measured in the dilution refrigerator. Figure S2a shows the temperature dependence of the resistance $R$ of the two samples. $R$ becomes as large as 10 M$\Omega$   at the lowest temperature, but both impedances are much smaller than the impedance of nanovoltmeter (10 G$\Omega$).
As shown in Fig. S2b, while the resistance is  different in the two samples, their resistivity is similar and in fair agreement with a previous report[S3].

The Seebeck coefficient, $S$, measured for both samples  are shown in Fig. S2c. As seen in the figure, there is a reasonable agreement between the two sets of the data. We conclude that  the low-temperature upturn of the Seebeck coefficient in the VRH regime leading to a finite value in the zero-temperature limit is reproduced in two samples. However, our data does not allow us to conclude if the magnitude of this residual term is sample-dependent or not.

\section{Comparison of the SDW anomaly with previous measurements}

Fig. S3 compares our thermopower data across the  SDW transition (T$_{SDW}\approx$ 12 K) with three earlier reports[S4,S5,S6]. As seen in the figure, qualitatively, the four set of results resolve a similar anomaly at T$_{SDW}$. But, the positive jump in thermopower has different amplitudes in the four set of data. The jump we resolved is much larger than those seen by the three other groups. The most plausible source for this discrepancy is  the magnitude of the applied thermal gradient. In our case we applied a temperature difference as  low as 0.2K which is only 1.5 percent of the absolute temperature. Usually, one applies temperature gradients larger than this, which would be fine when the Seebeck coefficient does not go through a drastic change with temperature. In this case, however, since the Seebeck coefficient is changing sign,  the application of a thermal different as small as 0.5 K (that is only 5 percent of the absolute temperature) would drastically squeeze the magnitude of the sign change in Seebeck coefficient. Since the magnitude of the thermal gradient is not indicated in earlier works, we cannot be sure that this is the origin of the discrepancy.

Let us note that there is  correlation between the magnitudes of Seebeck and specific heat anomalies. According to our data, the jump is $\Delta S \simeq$ 30 $\mu$VK$^{-1}$ for thermopower and $\Delta C \simeq$ 1.3 JK$^{-1}$mol$^{-1}$for specific heat[S7] . Multiplying $\Delta S$ by $eN_{Av}$ ($e$ is the electron charge and $N_{Av}$ is the Avogadro number), one find half of $\Delta C$. This in additional support for the accuracy of our data near T$_{SDW}$. We note also that this is a minor difference in face of the drastic enhancement of the Seebeck coefficient at low temperatures.

\begin{figure}
\includegraphics[width=8cm]{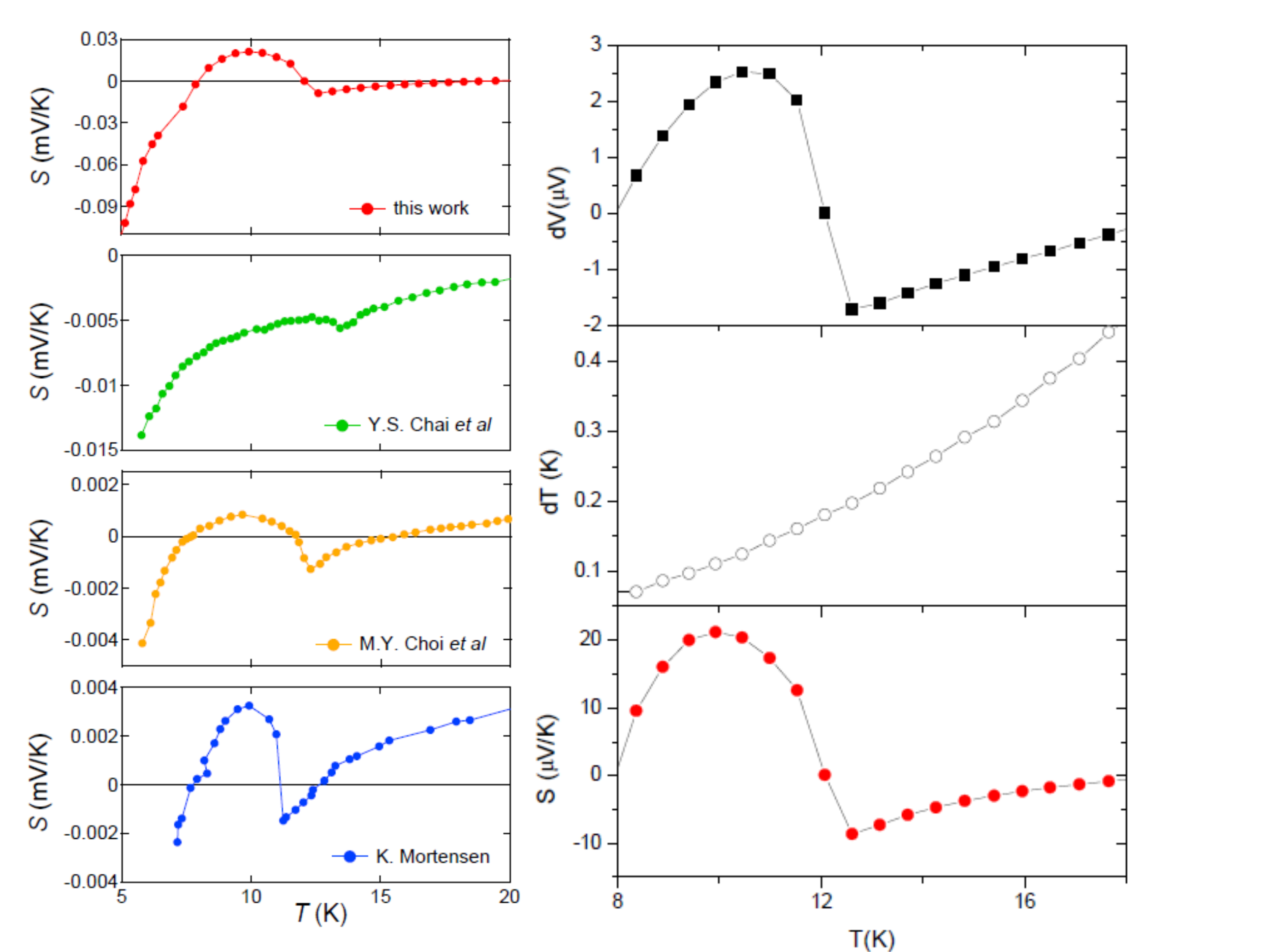}
\begin{flushleft}
{\bf Figure S3} Left: The temperature dependence of the Seebeck coefficient across the SDW transition according to our data and those reported by three earlier works. Right: The variation of the temperature difference, the voltage difference and the seebeck coefficient across the transition according to our data.
\end{flushleft}
\end{figure}

\begin{figure}
\includegraphics[width=7cm]{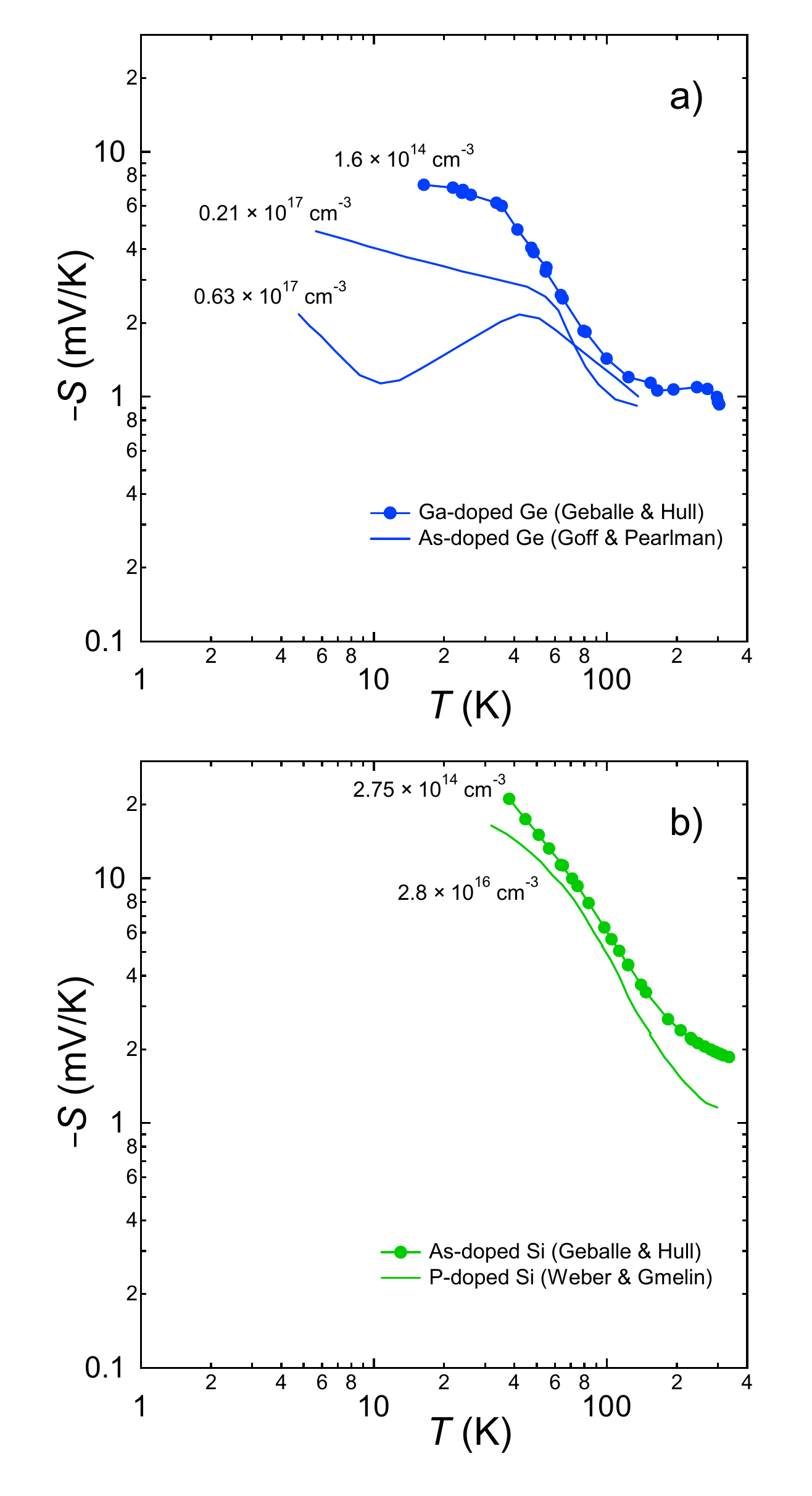}
\begin{flushleft}
{\bf Figure S4} The Seebeck coefficient of a) lightly doped germanium (adapted from Ref.~[S4,S5]) and b) lightly doped silicon (adapted from Ref.~[S6,S7]).
The carrier concentration of each sample is indicated in the figures.
\end{flushleft}
\end{figure}

\section{Available data on germanium and silicon}
Is our finding  generic to all semiconductors? The answer to this question is not known. The low-temperature thermoelectric response of semiconductors below the critical doping for metal-insulator transition has not been measured down to very low temperatures. This is true even in the case of the archetypical semiconductors, such as silicon and germanium. To the best of our knowledge, there are only a few studies devoted to them. In doped germanium with low arsenic concentrations, the Seebeck coefficient was reported to increase up to $\sim$ 5 mV/K with decreasing temperature down to 5 K in the regime where the resistivity shows a (VRH-like) behavior [S8]. Similar rapid increases of the Seebeck coefficient were also observed in barely doped germanium [S9] and silicon [S7,S8] while the lowest measured temperature was as high as $\sim$ 20 K. Fig. S3 compares the data for germanium [S8,S9] and silicon [S10,S11], respectively. As seen in the figures, in the least-doped sample, the Seebeck coefficient remains finite down to the lowest temperature investigated. However, further studies of the low-temperature Seebeck coefficient are required to see whether it survives in the zero temperature limit.

\noindent {[S1]} Y. Machida, K. Tomokuni, K. Izawa, G. Lapertot, G. Knebel, J.-P. Brison, and J. Flouquet, {\it Physical Review Letters} {\bf 109}, 156405 (2012).\\
\noindent {[S2]} S. Hartmann, N. Oeschler, C. Krellner, C. Geibel, S. Paschen, and F. Steglich, {\it Physical Review Letters} {\bf 104}, 096401 (2010).\\
\noindent {[S3]} Y. M. Kim, G. Mih$\acute{\rm a}$ly, H. W. Jiang, and G. Gr$\ddot{\rm u}$ner, {\it Synthetic Metals} {\bf 70}, 1287 (1995).\\
\noindent {[S4]} K. Mortensen, {\it Solid State Commun.} \textbf{44}, 643 (1982).\\
\noindent {[S5]} M.-Y. Choi, M. J. Burns, P. M. Chaikin E. M. Engler and R. I . Greene, {\it Phys. Rev. B}  \textbf{31}, 3576 (1985).\\
\noindent {[S6]} Y. S. Chai, H. S. Yang, J. Liu, C. H. Sun, H. X. Gao, X. D. Chen, L. Z. Cao and J. C. Lasjaunias, {\it Phys. Lett. A} \textbf{366}, 513 (2007).\\
\noindent {[S7]}D. K. Powell, K. P. Starkey, G. Shaw, Y. V. Sushko,  {\it Solid State Commun.} \textbf{119}, 637  (2001).\\
\noindent {[S8]} J. F. Goff and N. Pearlman, {\it Physical Review} {\bf 140}, 2151 (1965).\\
\noindent {[S9]} T. H. Geballe and G. W. Hull, {\it Physical Review} {\bf 94}, 1134 (1954).\\
\noindent {[S10]} T. H. Geballe and G. W. Hull, {\it Physical Review} {\bf 98}, 940 (1955).\\
\noindent {[S11]} L. Weber and E. Gmelin, {\it Applied Physics A} {\bf 53}, 136 (1991).\\

\begin{thebibliography}{}
\bibitem{callen1948} H. B. Callen, Phys. Rev. \textbf{73}, 1349 (1948)
\bibitem{macdonald1962}  D. K. C. Macdonald, Thermoelectricity, an introduction to the principles, John Wiley \& sons, New York (1962)
\bibitem{ziman1964}J. M. Ziman, Principles of the theory of Solids, Cambridge University Press, Cambridge (1964)
\bibitem{behnia2015} K. Behnia, Fundamentals of thermoelectricity, Oxford University Press, Oxford (2015)
\bibitem{zvyagin1973} I. P. Zvyagin,   Phys. Stat. Sol. (b) \textbf{68}, 443 (1973)
\bibitem{whall1981} T. E. Whall,J. Phys. C \textbf{14}, L887 (1981)
\bibitem{burns1985} M. J. Burns and P. M. Chaikin, J. Phys. C \textbf{18}, L743 (1985)
\bibitem{lien1999} N. Van Lien and D. Dinh Toi, Phys. Lett. A \textbf{261}, 108 (1999)
\bibitem{mahan2015} G.D. Mahan,  J. Electron. Mater. \textbf{44}, 431 (2015)
\bibitem{geballe1954} T. H. Geballe and G. W. Hull, Phys. Rev. \textbf{94}, 1134 (1954)
\bibitem{geballe1955} T. H. Geballe and G. W. Hull,  Phys. Rev. \textbf{98}, 940 (1955)
\bibitem{jerome1980} D. J\'erome, A. Mazaud, M. Ribault and K. Bechgaard,  J. Physique Lett. \textbf{41}, L 95 (1980)
\bibitem{jerome2012} D. J\'erome,  J. Supercond. Nov. Magn. 25, 633 (2012)
\bibitem{brown2015} S. Brown,  Physica C \textbf{514},  279 (2015)
\bibitem{chaikin1981}P. M. Chaikin, P. Haen, E. M. Engler, and R. L. Greene, Phys. Rev. B \textbf{24}, 7155 (1981)
\bibitem{kim1995} Y. M.  Kim, G.  Mih\'{a}ily, H. W. Jiang and G. Gr\"{u}ner,  Synthetic  Metals \textbf{70}, 1287 (1995)
\bibitem{petukhov2005} K. Petukhov and M. Dressel,  Phys. Rev. B \textbf{71}, 073101 (2005)
\bibitem{gruner1994} G. Gr\"{u}ner,  Rev. Mod. Phys. \textbf{66}, 1 (1994)
\bibitem{mott1968} N. F. Mott,  Journal of Non-Crystalline Solids \textbf{1},  1 (1968)
\bibitem{reviewVRH} B. I. Shklovskii and A. L. Efros, Electronic Properties of Doped Semiconductors, Springer-Verlag, Berlin (1984)
\bibitem{adam2014} O. Agam and I. L. Aleiner, Phys. Rev. B \textbf{89}, 224204 (2014)
\bibitem{mortensen1982} K. Mortensen,  Solid State Commun. \textbf{44}, 643 (1982)
\bibitem{choi1985} M.-Y. Choi, M. J. Burns, P. M. Chaikin E. M. Engler and R. I . Greene,  Phys. Rev. B \textbf{31}, 3576 (1985)
\bibitem{chai2007} Y. S. Chai, H. S. Yang, J. Liu, C. H. Sun, H. X. Gao, X. D. Chen, L. Z. Cao and J. C. Lasjaunias, Phys. Lett. A \textbf{366}, 513 (2007)
\bibitem{coroneus1993}  J. Coroneus, B. Alavi, and S. E. Brown,  Phys. Rev. Lett. \textbf{70}, 2332 (1993)
\bibitem{powell2001}  D. K. Powell, K. P. Starkey, G. Shaw, Y. V. Sushko, L. K. Montgomery, J. W. Brill, Solid State Commun. \textbf{119}, 637 (2001)
\bibitem{mortensen1983} K. Mortensen, E. M. Conwell and J. M. Fabre, Phys. Rev. B \textbf{28}, 5856 (1983)
\bibitem{mortensen1984} K. Mortensen, E. M. Engler,  Phys. Rev. B \textbf{29}, 842 (1984)
\bibitem{uji1997} S. Uji, J. S. Brooks, M. Chaparala, S. Takasaki, J. Yamada, and H. Anzai, Phys. Rev. B \textbf{55}, 12 446 (1997)
\bibitem{nagata2013} S. Nagata, M. Misawa, Y. Ihara, and A. Kawamoto, Phys. Rev. Lett. \textbf{110}, 167001 (2013)
\bibitem{barthel1993} E. Barthel, G. Quirion, P. Wzietek, D. J\'{e}rome, J. B. Christensen, M. J{\o}rgensen and K. Bechgaard, Europhys. Lett. \textbf{21}, 87 (1993)
\bibitem{izawa2007} K. Izawa, K. Behnia, Y. Matsuda, H. Shishido, R. Settai, Y. Onuki, and J. Flouquet, Phys. Rev. Lett. \textbf{99}, 147005 (2007)
\bibitem{machida2012} Y. Machida, K. Tomokuni, K. Izawa, G. Lapertot, G. Knebel, J.-P. Brison, and J. Flouquet,  Phys. Rev. Lett. {\bf 109}, 156405 (2012)
\bibitem{pourret2006}	A. Pourret, H. Aubin, J. Lesueur, C. A. Marrache-Kikuchi, L. Berg\'{e}, L. Dumoulin and K. Behnia,  Nature Phys. \textbf{2}, 683 (2006)
\bibitem{zhu2010}	Z. Zhu, H. Yang, B. Fauqu\'{e}, Y. Kopelevich and K. Behnia,   Nature Physics \textbf{6}, 26 (2010)
\bibitem {hartmann2010} S. Hartmann, N. Oeschler, C. Krellner, C. Geibel, S. Paschen and F. Steglich, Phys. Rev. Lett. \textbf{104}, 096401 (2010)
\bibitem{supplement}  See the supplemental material in appendix.
 \bibitem{efros1975} A. L. Efros and B. I. Shklovskii, J. Phys. C: Solid State Phys. \textbf{8}, L49(1975)
\bibitem{weber1991} L. Weber and E. Gmelin, Appl. Phys. A \textbf{53}, 136(1991)
\bibitem{fredrikse1953} H. P. R. Frederikse,  Phys. Rev. \textbf{92}, 248 (1953)
\bibitem{goff1965}J. F. Goff and N. Pearlman,  Phys. Rev. \textbf{140}, A2151 (1965)
\bibitem{ziegler1996} P. Ziegler, M. Lakner  and H. v. L\"{o}hneysen, Europhys. Lett. \textbf{33},  285 (1996)
\bibitem{heikes1961} R. R. Heikes  and R. W. Jr. Ure, Thermoelectricity: Science and Engineering, Interscience Publishers, New York, London (1961)
\bibitem{parrondo} J. M. R. Parrondo,	J. M. Horowitz	and T. Sagawa, Nature Physics \textbf{11}, 131 (2015); E. Lutz and S. Ciliberto, Phys. Today \textbf{68(9)}, 30 (2015)
\end{thebibliography}
\end{document}